\documentstyle[a4wide,12pt,epsf]{article}
\begin{document}
% TH FORMAT
\begin{flushright}
\baselineskip=12pt
{SUSX-TH-97/11}\\
{hep-th/9705244}\\
May 1997\\
%revised version
\end{flushright}

\begin{center}
%\vglue 0.5cm
{\LARGE \bf CP VIOLATION BY SOFT SUPERSYMMETRY BREAKING TERMS IN
ORBIFOLD COMPACTIFICATIONS \\}
\vglue 0.35cm
{D.BAILIN$^{\clubsuit}$ \footnote
{D.Bailin@sussex.ac.uk}, G. V. KRANIOTIS$^{\spadesuit}$ \footnote
 {G.Kraniotis@rhbnc.ac.uk} and A. LOVE$^{\spadesuit}$ \\}
%\vglue 0.2cm
	{$\clubsuit$ \it  Centre for Theoretical Physics, \\}
{\it University of Sussex,\\}
{\it Brighton BN1 9QJ, U.K. \\}
%\vgluw 0.2cm
{$\spadesuit$ \it  Department of Physics, \\}
{\it Royal Holloway and Bedford New College, \\}
{\it  University of London,Egham, \\}
{\it Surrey TW20-0EX, U.K. \\}
\baselineskip=12pt

\vglue 0.25cm
ABSTRACT
\end{center}

%\vglue 0.5cm
{\rightskip=3pc
\leftskip=3pc
\noindent
\baselineskip=20pt
The possibility of spontaneous breaking of $CP$ symmetry by the
expectation values of orbifold moduli is investigated  with particular
reference to $CP$ violating phases in soft supersymmetry breaking
terms. The effect of different mechanisms for stabilizing the
dilaton and the form of the non-perturbative superpotential on the
existence and size of these phases is studied. Non-perturbative 
superpotentials involving the absolute modular invariant $j(T)$,
such as may arise from $F-$ theory compactifications, are
considered.
 }

\vfill\eject
\setcounter{page}{1}
\pagestyle{plain}
\baselineskip=14pt
	
String theory may provide a new perspective on the longstanding
question of the origin of $CP$ violation. It has been argued
\cite{dine} that there is no explicit $CP$ symmetry breaking in
string theory whether perturbative or non-perturbative.
However, $CP$ violation might arise from complex expectation values
of moduli or other scalars \cite{dine}-\cite
{steve}. In all supergravity
theories, including those derived from string theory, there is the
possibility of $CP$ violating phases in the soft supersymmetry-
%\footnote{Any nonzero value for $d_n$,
%the neutron electric dipole
%moment, is an indication of $CP$ violation.
%In principle, complex soft superymmetry breaking terms in SUSY theories
%and their resulting phases
%can lead to large contributions to $d_n$. These  phases are constrained
%by experiment to be $\leq O(10^{-3})$. It is therefore a serious
%challenge for SUSY theories to explain why these phases are so small,
%i.e why soft susy breaking terms preserve $CP$ to such a high
%degree.
%As we will show, below, string supersymmetric theories relate the
%required smallness of $CP$ phases to properties of modular functions.}
breaking $A$ and $B$ terms and gaugino masses which are in addition
to a possible phase in the Kobayashi-Maskawa matrix and the 
$\theta$ parameter of QCD. (See for example, \cite{HALL} and references
therein.) In compactifications of string theory,
soft supersymmetry breaking terms can be functions of moduli
such as those associated with the radius and angles characterizing the
underlying torus of the orbifold compactification. Then, if
these moduli develop complex vacuum expectation values, this can
be fed through to the low energy supergravity as $CP$ 
violating phases.

Any nonzero value for $d_n$,
the neutron electric dipole
moment, is an indication of $CP$ violation.
In principle, complex soft superymmetry-breaking terms in SUSY theories
and their resulting phases
can lead to large contributions to $d_n$. These  phases are constrained
by experiment to be $\leq O(10^{-3})$. It is therefore a serious
challenge for SUSY theories to explain why these phases are so small,
i.e. why soft susy breaking terms preserve $CP$ to such a high
degree.
As we will show below, string supersymmetric theories relate the
required smallness of $CP$ phases to properties of modular functions.

To estimate the size of such $CP$ violating phases for orbifold
compactifications, it is first necessary to minimize the effective
potential to determine the expectation values of the $T$ moduli.
Such calculations may be sensitive to the solution proposed
to the problem of stabilizing the dilaton expectation value.
In an earlier paper \cite{steve} in the context of
orbifolds with broken $PSL(2,Z)$ modular symmetry and a single
gaugino condensate non-perturbative superpotential, no
assumption was made as to the mechanism for dilaton stabilization.
Instead, the dilaton expectation value $S$ and the corresponding
auxiliary $F_S$ were treated as free parameters but with
$Re S$ fixed to about 2, in line with the value of the inverse
gauge coupling constant squared at the string scale. Here,
we improve on such calculations by assuming that the dilaton 
is stabilized either by a multiple gaugino condensate 
\cite{DIXON},
or by stringy non-perturbative corrections to the dilaton
K$\rm{\ddot{a}}$hler potential. For simplicity, our treatment
here is restricted to a single overall modulus $T$ and unbroken
$PSL(2,Z)$ modular symmetry. Models with broken modular symmetries
will be discussed elsewhere \cite{US}. The calculations are
carried out both when the non-perturbative superpotential
has the generic form derived from gaugino condensation
\cite{font} including the Dedekind function $\eta(T)$, and when
the superpotential also involves \cite{miriam} the absolute
modular invariant $j(T)$. Non-perturbative superpotentials
involving $j(T)$ may, in principle, arise from orbifold
theories containing gauge non-singlet states which become
massless at some special values of the moduli \cite{miriam}
(though examples are lacking.) They may also arise from
$F-$ theory compactifications  \cite{witten}.

If we assume that the stabilization of the dilaton expectation
value at a minimum with a realistic value of Re$S$ is due to
a multiple gaugino condensate \cite{DIXON} (including hidden
sector matter), the general form of the non-perturbative
superpotential for a single overall modulus $T$ is of the
form
\begin{equation}
W_{np}=\sum_{a} h_{a}e^{(\frac{24\pi^2 S}{b_a})} (\eta(T))^
{-6(1-{\frac{8\pi^2}{b_a} {\tilde{\delta}}_{GS}})}
\label{multi}
\end{equation}
for some coefficients $h_a$, where the sum over 
$a$ is over factors in the hidden sector gauge group, the
$b_a$ are the corresponding renormalization group
coefficients, and the Green-Schwarz coefficient
$\tilde{\delta}_{GS}$ is normalized such that the 
dilaton and moduli dependent K$\rm{\ddot{a}}$hler potential is
\begin{equation}
K=-3\log(T+\bar{T})-\log y
\label{potk}
\end{equation}
with 
\begin{equation}
y=S+\bar{S}-\tilde{\delta}_{GS}\log(T+\bar{T})
\end{equation}

Since we do not wish to commit ourselves to any
particular choice of gaugino condensates nor of hidden sector
matter, we find it convenient to rewrite (1) in the
form
\begin{equation}
W_{np}=\Omega(\Sigma)\eta(T)^{-6}
\label{model}
\end{equation}
where 
\begin{equation}
\Sigma=S+2\tilde{\delta}_{GS} \log{\eta(T)}
\label{mod2}
\end{equation}
and
\begin{equation}
\Omega(\Sigma)=\sum_{a} h_{a} e^{\frac{24\pi^2}{b_a}\Sigma}
\label{mod3}
\end{equation}

In what follows, we shall treat $\Sigma$ as a parameter
to be chosen so that $y$ is approximately 4, and we shall
also treat
\begin{equation}
\rho\equiv \frac{\frac{d\Omega}{d\Sigma}}{\Omega}
\label{param}
\end{equation}
as a free parameter. The parameter $\rho$ is related 
to the dilaton auxiliary field $F_S$ by
\begin{equation}
\rho=\frac{1-F_S}{y}
\label{aux}
\end{equation}

If instead we assume that stabilization of the dilaton
expectation value is produced by stringy non-perturbative
corrections \cite{casa} to the dilaton K${\rm{\ddot{a}}}$hler
potential, then we write the dilaton and moduli dependent part
of the  K${\rm{\ddot{a}}}$hler potential as
\begin{equation}
K=-3 \log(T+\bar{T})+P(y)
\label{pymod}
\end{equation}
where $P(y)$ is a function to be determined by non-perturbative
string effects. In that case, we shall treat
$\frac{dP}{dy}$ and $\frac{d^2P}{dy^2}$, which we shall see
occur in the effective potential and the soft supersymmetry 
breaking terms, as free parameters.

The form of the effective potential which encompasses both
possibilities is
\begin{eqnarray}
V_{eff}&=& |W_{np}|^2\; e^{P(y)}\;(T+\bar{T})^{-3} \nonumber \\
       &\times& \Biggl[-3+\Bigl|\frac{dP}{dy}+\rho
\Bigr|^2\Bigl(\frac{d^2P}{dy^2}\Bigr)^{-1}+ \frac{|
\tilde{\delta}_{GS}\rho-3|^2}
{(3+\tilde{\delta}_{GS}\frac{dP}{dy})}
(T+\bar{T})^2|\hat{G}(T,\bar{T})|^2\Biggr]
\label{pote}
\end{eqnarray}
where
\begin{equation}
\hat{G}(T,\bar{T})=(T+\bar{T})^{-1}+ 2 \eta^{-1}\frac{d\eta}{dT}
\label{eisen}
\end{equation}
and $\rho$ has the value $\frac{24 \pi^2}{b}$ for the single
condensate case.
The soft supersymmetry-breaking terms may be calculated
by standard methods. (See for example \cite{ispa} and 
\cite{BEATRIZ}, from
which the earlier literature can be traced.)
The gaugino masses $M_a$ are given by
\begin{eqnarray}
M_a&=&m_{3/2}(Re f_a)^{-1}\times \Biggl[
\frac{\partial{\bar{f}_a}}{\partial{
\bar{S}}}\Bigl(\frac{d^2P}{dy^2}\Bigr)^{-1} \Bigl(\frac{dP}{dy}+\rho
\Bigr) \nonumber \\
&+&
(\frac{b_a^{'}}{8 \pi^2}-
\tilde{\delta}_{GS})\Bigl(1+\frac{\tilde{\delta}_{GS}\frac{dP}{dy}}{3}
\Bigr)^{-1}
\Bigl(\rho\frac{\tilde{\delta}_{GS}}{3}-1\Bigr)(T+\bar{T})^2|\hat{G}|^2\Biggr]
\label{gaugi}
\end{eqnarray}
where $b_a^{'}$ is the usual coefficient occuring in the
string loop threshold corrections to the gauge coupling
constant \cite{ispa,LOUIS}. Provided the dilaton
auxiliary field $F_S$ in (\ref{gaugi}) is real, there are
no $CP$ violating phases in the gaugino masses.
The soft supersymmetry-breaking terms are given by
\begin{eqnarray}
m_{3/2}^{-1}A_{\alpha\beta\gamma}&=&\Bigl(\frac{d^2P}{dy^2}
\Bigr)^{-1}\Bigl(
\frac{dP}{dy}+\rho\Bigr)\frac{dP}{dy} \nonumber \\
&+&\Bigl(1+\frac{\tilde{\delta}_{GS}}{3}\frac{dP}{dy}
\Bigr)^{-1}\Bigl(1-
\rho\frac{\tilde{\delta}_{GS}}{3}\Bigr)(T+\bar{T})\bar{\hat{G}}
\nonumber \\
&\times&
\Bigl(3+n_{\alpha}+n_{\beta}+n_{\gamma}-(T+\bar{T})\frac{\partial{
\log{h_{\alpha\beta\gamma}}}}{\partial{T}}\Bigr)
\label{asoft}
\end{eqnarray}
where the superpotential term for the Yukawa couplings of
$\phi_{\alpha},\phi_{\beta}$ and $\phi_{\gamma}$ is
$h_{\alpha\beta\gamma}\phi_{\alpha}\phi_{\beta}\phi_{\gamma}$,
the modular weights of these states are
$n_{\alpha},n_{\beta}$ and $n_{\gamma}$, and the usual
rescaling by a factor $\frac{W_{np}}{|W_{np}|}$ required
to get from the supergravity theory derived from
the orbifold compactification of the superstring
theory to the spontaneously broken globally supersymmetric
theory has been carried out. (See, for example, \cite{Iba:Spain}.)
The $\frac{\partial{\log{h_{\alpha\beta\gamma}}}}{\partial{T}}$
contribution to (\ref{asoft}) is essential for the modular invariance 
of $A_{\alpha\beta\gamma}$ and can make a significant  contribution
to any $CP$ violating phase. For illustrative purposes we have taken
$h_{\alpha\beta\gamma}$ to be of the form encountered 
\cite{Chun} when each of the
states $\phi_{\alpha},\phi_{\beta}$ and $\phi_{\gamma}$ is in the
particular twisted sector of the $Z_{3} \times Z_{6}$ orbifold with the
same twisted boundary conditions as the twisted sector of the
$Z_3$ orbifold.
This is an appropriate choice because the $Z_3\times Z_6$ orbifold
has three $N=2$ moduli,$\;T_i,\;i=1,2,3$, so that the model of a single
overall modulus $T=T_1=T_2=T_3$ is consistent. In this case, if we
arrange $h_{\alpha\beta\gamma}$ to be covariant under the $T\rightarrow
\frac{1}{T}$ modular transformation, it is a product of
3 factors, one for each complex plane,of the form
\begin{equation}
h(T_i,k_i=0)+(\pm\sqrt{3}-1)h(T_i,k_i=1)
\end{equation}
where
\begin{equation}
h(T_i,k_i)\sim e^{-\frac{2}{3}\pi k_i^2 T_i}\Bigl[\Theta_3(ik_iT_i,
2i T_i)\Theta_3(ik_iT_i,6iT_i)+
\Theta_2(ik_iT_i,2iT_i)\Theta_2(ik_iT_i,6iT_i)\Bigr]
\label{japan}
\end{equation}
Each of the modular weights $n_{\alpha},n_{\beta}$ and $n_{\gamma}$
 has the value -2.
%We have chosen the plus sign in (\ref{japan}) for each complex plane.

The expression for the soft supersymmetry-breaking $B$ term depends
on the mechanism adopted for generating the $\mu$ term
for the Higgs scalars $H_1$ and $H_2$, with corresponding
superfields $\phi_1$ and $\phi_2$. If we assume that the
$\mu$ term is generated non-perturbatively as an
explicit superpotential term $\mu_W\phi_1\phi_2$, then the
$B$ term, which in this case we denote by $B_W$, is given
by
\begin{eqnarray}
m_{3/2}^{-1}B_W&=&-1+\Bigl(\frac{d^2P}{dy^2}\Bigr)^{-1}\Bigl(\frac{dP}{dy}+
\bar{\rho}\Bigr)
\Bigl(\frac{dP}{dy}+\frac{\partial{\log\mu_W}}{\partial{S}}\Bigr) \nonumber \\
&+&\Bigl(1+\frac{\tilde{\delta}_{GS}}{3}\frac{dP}{dy}\Bigr)^{-1}\Bigl(
1-\bar{\rho}\frac{\tilde{\delta}_{GS}}{3}\Bigr)(T+\bar{T})\bar{\hat{G}}
\nonumber \\
&\times&\Bigl(3+n_1+n_2-(T+\bar{T})\frac{\partial{\log\mu_W}}{
\partial{T}}-\tilde{\delta}_{GS}\frac{\partial{\log\mu_W}}
{\partial{S}}\Bigr)
\label{bsoft}
\end{eqnarray}
where $n_1$ and $n_2$ are the modular weights of the
Higgs scalar superfields $\phi_1$ and $\phi_2$, and again
the appropriate rescaling has been carried out.

On the other hand, if the $\mu$ term is generated by a
term of the form $Z\phi_1\phi_2$+h.c. in the
K$\rm{\ddot{a}}$hler potential mixing the 
Higgs superfields \cite{anto}, then
(before rescaling the Lagrangian) the $B$ term, which
we denote by $B_Z$ in this case, is given by
\begin{eqnarray}
-m_{3/2}^{-1}\mu_{Z}^{eff}B_Z&=&W_{np}Z\times\Biggl[2+\Bigl(\frac{T+
\bar{T}}{3+\tilde{\delta}_{GS}\frac{dP}{dy}}(\tilde{\delta}_{GS}
\rho-3)\hat{G}(T,\bar{T})+\rm{h.c.}\Bigr)\Biggr] \nonumber \\
&+&W_{np} Z\times\Bigl[-3+\Bigl|\frac{dP}{dy}+\rho\Bigr|^2
\Bigl(\frac{d^2P}{
dy^2}\Bigr)^{-1}  \nonumber \\
&+&\frac{|\tilde{\delta}_{GS}\rho-3|^2(T+\bar{T})^2}
{(3+\tilde{\delta}_{GS}\frac{dP}{dy})}|\hat{G}(T,\bar{T})|^2\Bigr]
\label{bka}
\end{eqnarray}
and the effective $\mu$ term in the superpotential has
$\mu=\mu_{Z}^{eff}$ where
\begin{equation}
\mu_{Z}^{eff}=|W_{np}|Z\Bigl(1-\frac{\tilde{\delta}_{GS}}{3}\rho+
\frac{T+\bar{T}}{3}(\tilde{\delta}_{GS}\rho-3)\hat{G}(T,\bar{T})\Bigr)
\label{mu}
\end{equation}
To obtain the final form for $B_Z$ in the low energy supersymmetry
theory, rescaling of the Lagrangian by $\frac{W_{np}}{|W_{np}|}$
has to be carried out. In that case, any $CP$ violating phase
derives from (\ref{mu}). This mechanism requires \cite{anto}
that the Higgs scalars are in the untwisted sector of the 
orbifold and are associated with the $T$ and $U$ modulus
for a complex plane on which the point group acts as $Z_2$.
In the spirit of retaining only a single overall $T$ modulus
the auxiliary field of the $U$ modulus has been set to zero
in deriving (\ref{bka}) ans (\ref{mu}), which is equivalent
to assuming that the supersymmetry breaking is dominated by
the dilaton and the $T$ modulus.
 
In the case of multiple gaugino condensate with perturbative
K$\rm{\ddot{a}}$hler potential, minimization of the
effective potential at fixed $\Sigma$ for different
real values of the parameter $\rho$ with Re S taken to
be about 2 leads to the following conclusions.
It can be seen analytically that the fixed points 
of $PSL(2,Z)$ at $T=1$ and $T=e^{\frac{i\pi}{6}}$, at
which $\hat{G}(T,\bar{T})$ is zero, are always
extrema  (even for $\tilde{\delta}_{GS}\neq 0$.)
For $0.1\leq\rho\leq 0.4$, the minimum is at a real value
of $T$ which approaches 1 as $\rho$ approaches 0.42.
For $0.42 \leq \rho \leq 0.75$, $T$ remains at the fixed 
point at $T=1$, and for $\rho\geq 0.8$ the minimum is at
the other fixed (See fig.1 ) point at $T=e^{\frac{i\pi}{6}}$. (There
are of course also minima at points obtained from
these minima by modular transformations.) This resembles
what happens for a single condensate but treating
the dilaton auxiliary field $F_S$ as a free parameter
to simulate dynamics stabilizing the dilaton expectation
value \cite{ferrara}.

Gaugino condensate models (with perturbative 
K$\rm{\ddot{a}}$hler potential) in general have 
negative vacuum energy at the minimum. However, as other
authors have emphasized \cite{ferrara}, the solution 
to the vanishing cosmological constant problem is probably
in the realm of quantum gravity and, as the present type
of discussion treats gravity classically, we need not
necessarily impose vanishing vacuum energy as a constraint
on the theory. On the other hand, if we do arrange for
zero vacuum energy by introducing an extra matter field
which does not mix with the dilaton and moduli
fields \cite{Iba:Spain}, then the effect in the minimization
of the effective potential with respect to $T$ is that the
factor premultiplying the bracket in (\ref{pote}) is not
to be differentiated.
Then, for $0.25\leq \rho \leq 2.15$ minima occur
at the fixed points  at $T=1$ and $T=e^{\frac{i \pi}{6}}$.
For $\rho \geq 2.2$ there is a single real minimum.

In the case of a single gaugino condensate, but with
the dilaton expectation value being stabilized by
stringy non-perturbative corrections to the dilaton
K$\rm{\ddot{a}}$hler potential, minimization of the
effective potential with $y$ fixed at 4 for different
values of the parameters $\frac{dP}{dy}$ and 
$\frac{d^2P}{dy^2}$ leads instead to the following
outcome. For a wide range of choices of these parameters
$T$ is either at a fixed point value or it is real.
However, for some choices of the parameters $T$ takes complex
values which differ from the fixed point value $e^{\frac{i\pi}{6}}$.
For example, for $\frac{dP}{dy}=-11/4$ and 
$\frac{d^2P}{dy^2}=-1.3$, one obtains
\begin{equation}
T|_{min}=5.234339+0.0009575 i
\label{parad}
\end{equation}
and the potential is very flat.

The consequences of these  values of the modulus $T$ at the
minimum for possible $CP$ violating phases in the soft
supersymmetry-breaking terms are rather striking.
It might have been thought {\it a priori} that when
$T$ is at a fixed point at $e^{\frac{i \pi}{6}}$
a CP violating phase of order $10^{-1}$ might be induced.
However, as has been observed earlier \cite{steve}, if
$T$ is precisely at a fixed point value, and so at a zero
of $\hat{G}(T,\bar{T})$, the $CP$ violating phase
vanishes identically, as can be seen from (\ref{asoft})-
(\ref{mu}). 
Thus, for the case where the dilaton is stabilized
by a multiple gaugino condensate with perturbative 
K$\rm{\ddot{a}}$hler potential, there are no
$CP$ violating phases in the soft supersymmetry breaking
terms. In the case of a single gaugino condensate
with the dilaton stabilized by non-perturbative
corrections to the dilaton K$\rm{\ddot{a}}$hler potential,
the conclusion is the same for a wide range of values
of $\frac{dP}{dy}$ and $\frac{d^2P}{dy^2}$.
However, in this case, it is possible at the minimum
for $T$ to be at a complex value away from the fixed
point as, for example, in (\ref{parad}). At first sight,
it appears that there might then be a $CP$ violating phase of order
$10^{-4}$.
However, the $CP$ violating phases are far smaller than this
(of order $10^{-15}$ for $T$ as in (\ref{parad}).)
The reason for this is the very rapid variation of the
imaginary part of $\hat{G}(T,\bar{T})$ with $Re T$ 
%Also is worth noticing (since we obtain
%real solutions not at the fixed points) 
%that the imaginary part
%of $\hat{G}(T,\bar{T})$ decreases rapidly with Re $T$ 
as Re $T$ moves away from 1 if Im $T$ is held fixed. The imaginary 
part of 
$\hat{G}(T,\bar{T})$ varies by 11  orders of
magnitude as Re $T$ goes from $\frac{\sqrt{3}}{2}$
to 5.0 (See fig.2 ). The possibility of suppressing $CP$ violating
phases in this way has been suggested earlier
\cite{steve} in the context of orbifold models
with broken $PSL(2,Z)$ modular symmetries.

If, as discussed in the introduction, we allow the possibility
that the non-perturbative superpotential $W_{np}$ may involve
the absolute modular invariant $j(T)$, as well as the 
Dedekind eta function \cite{miriam}, then the situation
is very different. Then $W_{np}$  contains an extra factor
$H(T)$ where the most general form of $H(T)$ to avoid
singularities inside the fundamental domain \cite{miriam} is
\begin{equation}
H(T)=(j-1728)^{m/2}j^{n/3} P(j)
\label{soti}
\end{equation}
 where $m$ and $n$ are integers and $P(j)$ is a polynomial
in $j$. This results in modification of
(\ref{pote}) and (\ref{asoft})-(\ref{mu}) by the
replacement of $(\tilde{\delta}_{GS} \;\rho-3)\hat{G}(T,\bar{T})$
by $(\tilde{\delta}_{GS}\; \rho-3)\hat{G}(T,\bar{T})+
\frac{dln H}{dT}$. It is then possible, for some choices
of $H$, to obtain (complex) minima of the effective potential
for $T$ that lead to $CP$ violating phases in the
soft supersymmetry breaking terms of order $10^{-4}-10^{-1}$.
Let us start with the case of stabilizing the dilaton
by multiple gaugino condensate, then for $P(j)=1$ and $m=n=1$,
$\tilde{\delta}_{GS}=-\frac{30}{8 \pi^2}, \; \rho=0.45$, we find 
that the minimum is on the unit circle at
\begin{equation}
T|_{min}=0.971353713\pm 0.2376383050 i
\end{equation}
For the Yukawa couplings which 
we have considered (see below), this  leads to a $CP$ violating phase
not greater than $10^{-4}$. 
%For the same values for $P(j),m,n,
%\tilde{\delta}_{GS}$ but for $\rho=0.8$ leads to
%a phase for the $A$ term of $O(10^{-3})$.
Also for $P(j)=1$ and $m=n=1$,
$\tilde{\delta}_{GS}=-\frac{50} {8 \pi^2}, \; \rho=0.26$ the minimum is 
also on the unit circle at
\begin{equation}
T|_{min}=0.971352323\pm 0.237643985 i
\end{equation}
which again leads to $CP$ violating phase not greater
than $10^{-4}$ in the $A$ term.
% of $4\times 10^{-4}$.
%For $\rho\geq 0.8$ in this case we obtain $\phi(A)\geq 3\times 10^{-3}$.
%\footnote{
%The Yukawa couplins associated with twisted sector matter fields
%transform as modular forms as follows under $SL(2;Z)$:
%$h_{\alpha\beta\gamma}(T)\rightarrow\;
%h_{\alpha\beta\gamma}(T)(ic T+d)^{-3-n_{\alpha}-
%n_{\beta}-n_{\gamma}}$. A modular form of weight $r$ regular in the
%fundamental domain of the $T-$modulus is generically given by:
%$F(T)=\eta(T)^{2r}(\frac{G_{6}(T)}{\eta(T)^{12}})^{m}(\frac{G_{4}(T)}
%{\eta(T)^{8}})^{n}$, with $G_{6}(T)$and $G_{4}(T)$ the Eiseinstein series
%with modular weights 6 and 4 respectively.} 
On the other hand, if we assume that the dilaton is stabilized by 
non-perturbative corrections to the K$\rm{\ddot a}$hler potential,
then it is possible to find minima of the effective potential at
complex values of the $T-$ modulus not only on the boundary of
the fundamental domain {\it but also inside the fundamental domain}.
In this case larger $CP$ violating phases arise.
Indeed, it is possible for some values of the parameters
$\frac{dP}{dy}$ and $\frac{d^2P}{dy^2}$ to obtain phases that exceed
the current experimental limit. As a result we can constrain our
non-perturbative parameter space.
For instance, for $\frac{dP}{dy}=-1.5$ and 
$\frac{d^2P}{dy^2}=-0.2,\;m=1,n=3,\delta_{GS}=-30$ we obtain
the following solutions:
\begin{equation}
T|_{min}=1.01196232+ 0.16800043\;i
\label{fird}
\end{equation}
and its $T-dual$ under the generator $T\rightarrow \frac{1}{T}$,
\begin{equation}
T|_{min}=0.96167455-0.15965193\;i
\label{secd}
\end{equation}
Both minima lead to a phase $\phi(A)$ of order $10^{-2}$.
The foregoing results need a little amplification. 
For minima connected by $T\rightarrow T+i$ 
the Yukawa $h_{\alpha\beta\gamma}=h(T,k=0)$ leads 
to the same $CP$ violating phases at both minima, while 
for minima connected by $T\rightarrow\frac{1}{T}$ the 
Yukawa $h_{\alpha\beta\gamma}=
h(T,k=0)+(\sqrt{3}-1)h(T,k=1)$, which 
transforms as $h_{\alpha\beta\gamma}(1/T)=Th_{\alpha\beta\gamma}(T)$, also 
leads to the same $CP$ violating phases
at both minima. Both are of order $10^{-2}$.
However, since there is no linear combination of 
Yukawas which has modular weight 1
with respect to {\it all} modular transformations, we cannot do
better than characterize the scale of the CP violating  phases in this
way. It is important to note that since $V_{eff}$ is modular invariant,
the calculation of the electric dipole moment of the neutron,
for example, will necessarily yield a modular invariant
result, presumably 
with magnitude characteristic of the order  
$10^{-2}$ scale of the $CP$ violating phase of $A_{\alpha\beta\gamma}$.
This calculation will necessarily entail contributions from more than
one $A$ term. 

Similarly, for $\frac{dP}{dy}=-1.4,\;\frac{d^2P}{dy^2}=-0.1,\;m=1,n=3,
\delta_{GS}=-30$, see fig.3, we obtain
\begin{equation}
T|_{min}=0.79314323+ 0.11307387\;i
\label{t1}
\end{equation}
its $T-dual$ under $T\rightarrow \frac{1}{T}$
\begin{equation}
T|_{min}=1.23569142-0.17616545\;i
\label{t2}
\end{equation}
as well as the $T-dual$ of the later under $T\rightarrow T+i$
\begin{equation}
T|_{min}=1.23569142+0.82383457\;i
\label{t3}
\end{equation}
At the above 3-points of the moduli space we obtain a phase
$\phi(A)$ of order $10^{-3}-10^{-2}$.

In conclusion, whether the dilaton expectation value is
stabilized by a multiple gaugino condensate or by
stringy corrections to the dilaton 
K$\rm{\ddot{a}}$hler potential, we have found that,
provided the superpotential does not contain the
absolute modular invariant $j(T)$, the
$CP$ violating phases
in the soft supersymmetry-breaking
terms are either zero
or much smaller than $10^{-3}$.
Zero phases occur when the minimum
for the modulus $T$ is at a zero of $\hat{G}(T,\bar{T})$ or
is real.
Phases much smaller than $10^{-3}$ occur when $T$ is at a complex
value with the real part of $T$ far from its value
at a zero of $\hat{G}(T,\bar{T})$, because of the rapid
variation of the imaginary part of  $\hat{G}(T,\bar{T})$
as  $Re\;T$ varies. However, if we
allow the more general possibility that $W_{np}$
involves $j(T)$
as well as $\eta(T)$, as may arise from orbifold theories
if the theory contains gauge non-singlet states that
are zero at some special values of the moduli
\cite{miriam}, or may arise from $F-$ theory compactifications
\cite{witten}, then it is possible in some models to obtain
$CP$ violating phases in the soft supersymmetry-breaking terms
of order $10^{-4}-10^{-1}$. The largest phases occur for 
 minima of the potential inside
the fundamental domain of the $PSL(2,Z)$ $T-$ modulus.
\section*{Acknowledgements}
This research is supported in part by PPARC.

\newpage
\begin{figure}
\epsfxsize=6in
\epsfysize=8in
\epsffile{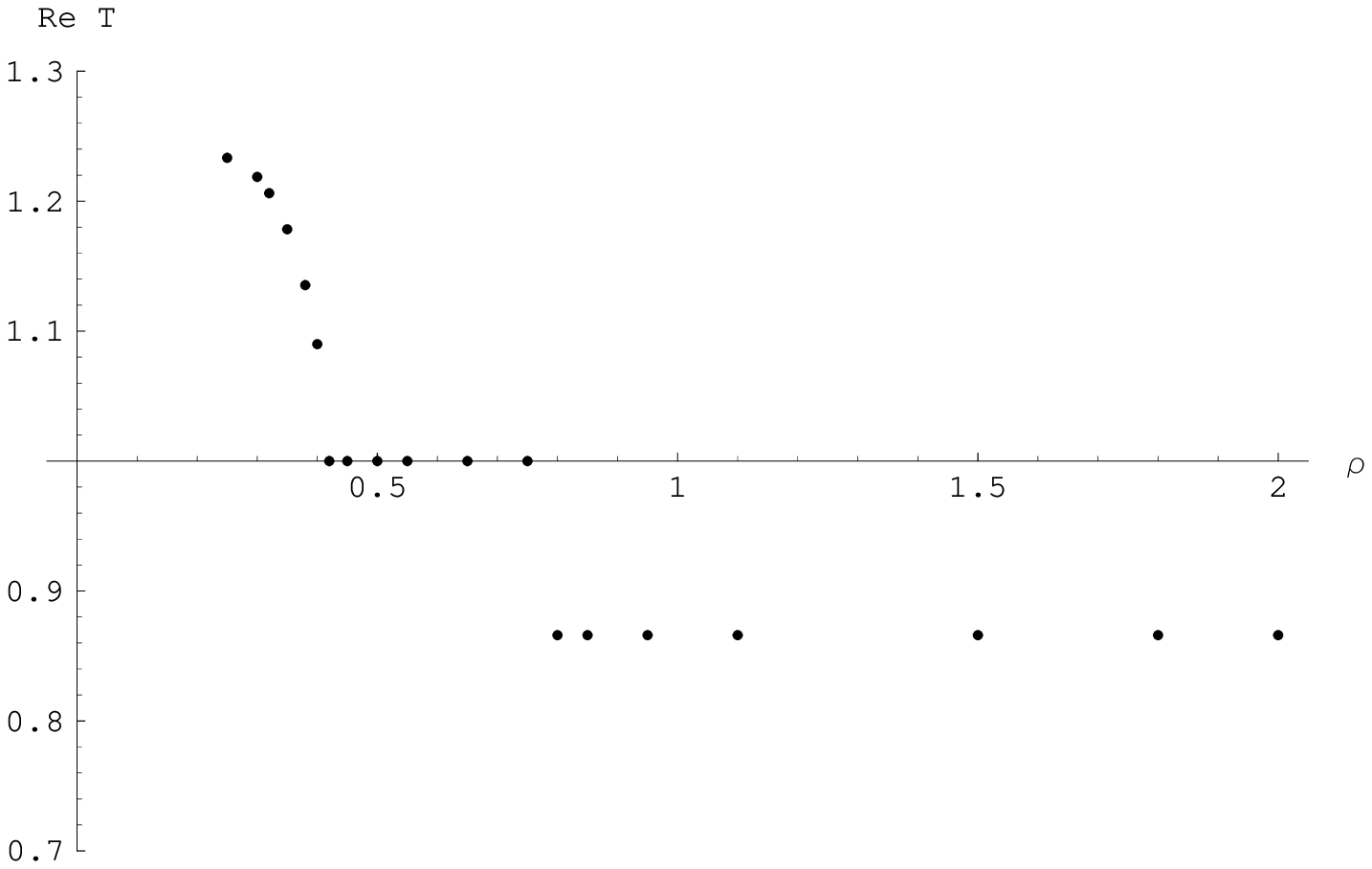}
%\hspace*{-2cm} \special{sp.ps}
%\vspace*{30cm}
\caption{Variation of $Re T$ at the minimum of $V_{eff}$ with
$\rho$, defined in eqn.(8), in multiple gaugino condensate models.}
\end{figure}
\newpage
\begin{figure}
\epsfxsize=6in
\epsfysize=4in
\epsffile{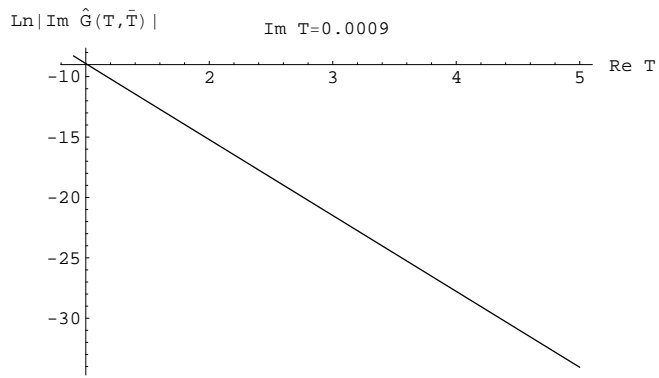}
%\vspace*{15cm}
%\hspace*{0.1cm} \special{run9.ps}
%\vspace*{1cm}
\caption{Variation of the logarithm of $Im \hat{G}$ with respect to $Re T$.}
\end{figure}
\newpage
\begin{figure}
\epsfxsize=6in
\epsfysize=6in
\epsffile{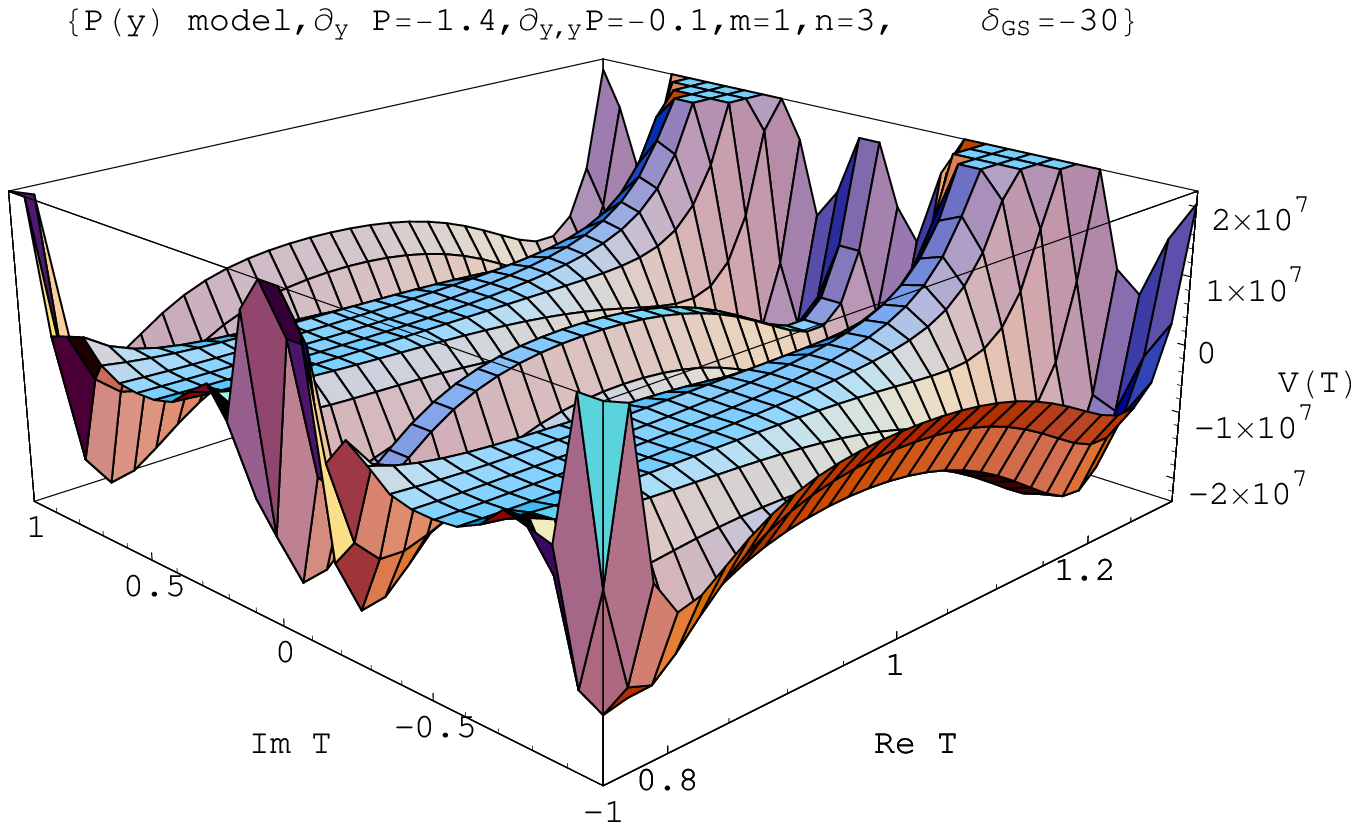}
%\vspace*{20cm}
%\hspace*{-2cm} \special{spa.ps}
%\vspace*{15cm}
\caption{The effective potential for the non-perturbative
dilaton K$\rm{\ddot{a}}$hler potential with the parameters as defined
in the text.}
\end{figure}


\begin{thebibliography}{99}
\bibitem{dine} M. Dine, R.G. Leigh and D.A. MacIntire, Phys.Rev.Lett.
69(1992),2030
\bibitem{IB:LU} L.E. Ib$\rm{\acute{a}\tilde{n}}$ez and D.L$\rm{\ddot{u}}$st, Phys.Lett.B267
(1991) 51
\bibitem{Iba:Spain} A. Brignole, L.E. Ib$\rm{\acute{a}\tilde{n}}$ez and C.Mu$\rm{\tilde{n}}$oz,
Nucl. Phys. B422(1994)125
\bibitem{steve} B. Acharya, D. Bailin, A. Love, W.A. Sabra and
S. Thomas, Phys.Lett. B357(1995)387
\bibitem{HALL} M.Dugan, B.Grinstein and L.Hall, Nucl. Phys. B255
(1985)413
\bibitem{DIXON}
J.A. Casas, Z. Lalak, C. Mu$\rm{\tilde{n}}$oz and G.G. Ross, Nucl. Phys. B347(1990)243;
N.V. Krasnikov, Phys. Lett. B193(1987)37;
T.R. Taylor, Phys.Lett.B252(1990)59
\bibitem{BANK} T.Banks and M.Dine, Phys.Rev. D50(1994)7454;
S.H. Shenker, Proceedings of Cargese Workshop on Random Surfaces,
Quantum Gravity and Strings, Cargese 1990.
\bibitem{casa} J.A. Casas, Preprint SCIPP--96--20,IEM-FT-129/96
\bibitem{US} D. Bailin, G.V. Kraniotis and A. Love, to be
submitted to Nucl. Phys. B.
\bibitem{font} A. Font, L.E. Ib$\rm{\acute{a}\tilde{n}}$ez, D. L$\rm{\ddot{u}}$st and
F. Quevedo, Phys.Lett.B245(1990) 401;
S. Ferrara, N. Magnoli, T.R. Taylor and G. Veneziano,
Phys.Lett.B245(1990) 409;
P. Binetruy and M.K. Gaillard, Phys.Lett.B253(1991)119
\bibitem{miriam} M. Cveti$\rm{\check{c}}$, A. Font, L.E. Ib$\rm{\acute{a}\tilde{n}}$ez, 
D. L$\rm{\ddot{u}}$st and F. Quevedo, Nucl.Phys.B361(1991)194
\bibitem{witten} R. Donagi, A. Grassi and E. Witten, hep-th/9607091;
G. Curio and D. L$\rm{\ddot{u}}$st, hep-th/9703007.
\bibitem{ispa} L.E. Ib$\rm{\acute{a}\tilde{n}}$ez and D. L$\rm{\ddot{u}}$st, Nucl.Phys.B382
(1992)305
\bibitem{BEATRIZ} B. de Carlos, J.A. Casas and C. Mu$\rm{\tilde{n}}$oz, Nucl.Phys.B399
(1993)623
\bibitem{LOUIS} L.J. Dixon, V.S. Kaplunovsky and J.Louis, Nucl.Phys.B355
(1991)649;
J.P. Derendinger, S. Ferrara, C.Kounnas and F. Zwirner, Nucl.Phys. B372
(1992) 145.
\bibitem{Chun}E.J. Chun, J. Mas, J. Lauer and H.P. Nilles, Phys.Lett.B233(1989)141; D. Bailin, A. Love and W.A. Sambra, Nucl. Phys. B403 (1993) 265
\bibitem{anto} I. Antoniadis, E.Gava, K.S. Narain and T.R. Taylor,
Nucl. Phys. B432(1994) 187
\bibitem{ferrara} S.Ferrara, N. Magnoli, T.R. Taylor and G. Veneziano,
Phys.Lett.B245(1990)409;
D. L$\rm{\ddot{u}}$st and T.R. Taylor, Phys.Lett.B253(1991)335
\end{thebibliography}
\end{document}